\documentclass[conference]{IEEEtran}%
\usepackage{amsfonts}
\usepackage{amsmath}
\usepackage{amssymb}
\usepackage{graphicx}
\usepackage{multicol}
\usepackage{xcolor}
\usepackage{pgf}%
\newtheorem{theorem}{Theorem}

\newtheorem{corollary}[theorem]{Corollary}

\newtheorem{definition}[theorem]{Definition}

\newtheorem{lemma}[theorem]{Lemma}

\newtheorem{proposition}[theorem]{Proposition}

\DeclareMathOperator*{\esssup}{ess\,sup}
\begin{document}

\title{R\'{e}nyi Divergence and Majorization}

%

%TCIMACRO{\TeXButton{Author Information}{\author{
%\authorblockN{Tim van Erven}
%\authorblockA{Centrum Wiskunde \& Informatica\\
%Amsterdam, The Netherlands\\
%E-mail: Tim.van.Erven@cwi.nl}
%\and\authorblockN{Peter Harremo\"es}
%\authorblockA{Copenhagen Business College\\
%Copenhagen, Denmark\\
%E-mail: harremoes@ieee.org}
%}}}%
%BeginExpansion
\author{
\authorblockN{Tim van Erven}
\authorblockA{Centrum Wiskunde \& Informatica\\
Amsterdam, The Netherlands\\
E-mail: Tim.van.Erven@cwi.nl}
\and\authorblockN{Peter Harremo\"es}
\authorblockA{Copenhagen Business College\\
Copenhagen, Denmark\\
E-mail: harremoes@ieee.org}
}%
%EndExpansion
%

%TCIMACRO{\TeXButton{Make Title}{\maketitle}}%
%BeginExpansion
\maketitle
%EndExpansion

%* For limit of decreasing sigma-algebras:
%add condition that 0 <= alpha < 1 or D_\alpha < \infty.
%

%TCIMACRO{\TeXButton{Begin abstract}{\begin{abstract}}}%
%BeginExpansion
\begin{abstract}%
%EndExpansion

R\'{e}nyi divergence is related to R\'{e}nyi entropy much like information
divergence (also called Kullback-Leibler divergence or relative entropy) is
related to Shannon's entropy, and comes up in many settings. It was introduced
by R\'{e}nyi as a measure of information that satisfies almost the same axioms
as information divergence.

We review the most important properties of R\'{e}nyi divergence, including its
relation to some other distances. We show how R\'{e}nyi divergence appears
when the theory of majorization is generalized from the finite to the
continuous setting. Finally, R\'{e}nyi divergence plays a role in analyzing
the number of binary questions required to guess the values of a sequence of
random variables.%

%TCIMACRO{\TeXButton{End abstract}{\end{abstract}}}%
%BeginExpansion
\end{abstract}%
%EndExpansion

\sloppy

\section{Introduction}

Since Shannon's introduction of his entropy function various other similar
measures of uncertainty or information have been introduced. Most of these
have found no applications and some have found applications only in quite
special cases. An exception is formed by R\'{e}nyi entropy and R\'{e}nyi
divergence, which pop up again and again. They are far from being as well
understood as Shannon entropy and Shannon divergence, and do not have as
simple an interpretation. Erdal Arikan observed that the discrete version of
R\'{e}nyi entropy is related to so-called guessing moments \cite{Arikan1996}.

In this short note we shall first review the most important properties of
R\'{e}nyi divergence in Section~\ref{sec:divergence}. In Section
\ref{sec:majorization} we give a very brief introduction to Markov ordering
and its relation to majorization. Then in
Sections~\ref{sec:DivergenceMajorization}, and \ref{sec:continuity} we relate
R\'{e}nyi divergence to the theory of majorization. And finally, in
Section~\ref{sec:guessing} we will show that, like its entropy counterpart,
R\'{e}nyi divergence is related to guessing moments.

\section{R\'{e}nyi Divergence\label{sec:divergence}}

Let $P$ and $Q$ be probability measures on a measurable space $(\mathcal{X}%
,\mathcal{F})$, and let $p$ and $q$ be their densities with respect to a
common $\sigma$-finite dominating measure $\mu$. Then for any $0<\alpha
<\infty$ except $\alpha=1$, the \emph{R\'{e}nyi divergence} $D_{\alpha}$ of
\emph{order} $\alpha$ of $P$ from $Q$ is defined as
\begin{equation}
D_{\alpha}(P\Vert Q)=\frac{1}{\alpha-1}\log\int p^{\alpha}q^{1-\alpha
}\ \mathrm{d}\mu, \label{eqn:commondefinition}%
\end{equation}
with the conventions that $p^{\alpha}q^{1-\alpha}=0$ if $p=q=0$, even for
$\alpha<0$ and $\alpha>1$, and that $x/0=\infty$ for $x>0$. Continuity
considerations lead to the following extensions for $\alpha\in\{0,1\}$:
\begin{align*}
D_{0}(P\Vert Q)  &  =\lim_{\alpha\downarrow0}D_{\alpha}(P\Vert Q)=-\log
Q(p>0),\\
D_{1}(P\Vert Q)  &  =\lim_{\alpha\uparrow1}D_{\alpha}(P\Vert Q)=D(P\Vert Q),
\end{align*}
where $D(P\Vert Q)=\int p\log p/q\ \mathrm{d}\mu$ (with the conventions that
$0\log0/x=0$ and $x\log x/0=\infty$ if $x>0$) denotes the \emph{information
divergence}, which is also known as Kullback-Leibler divergence or relative
entropy. For $\alpha>0$, it was introduced by R\'{e}nyi \cite{Renyi1961}, who
provided an axiomatic characterization in terms of \textquotedblleft
intuitively evident postulates\textquotedblright. An operational
characterizations of R\'{e}nyi divergence via coding has been described
\cite{Harremoes2006d}.

We will first review some of the basic properties of $D_{\alpha}$. Whenever
these properties can easily be derived from known results, we will point to
the relevant literature. For other properties, space requirements limit us to
only hint at their proofs. A longer version of this paper with full proofs
will be published elsewhere, and will include results for negative values of
the order $\alpha$.

Let us start by noting that, for finite orders $0<\alpha\neq1$, $D_{\alpha}$
is a continuous, strictly increasing function of the power divergence%
\[
d_{\alpha}(P,Q)=\frac{\int p^{\alpha}q^{1-\alpha}\ \mathrm{d}\mu-1}{\alpha
-1}.
\]
As $d_{\alpha}$ are \emph{$f$-divergences}, we may derive properties for
$D_{\alpha}$ from general properties of $f$-divergences \cite{Liese2006}.

In particular, R\'{e}nyi divergence satisfies the \emph{data processing
inequality}
\[
D_{\alpha}(P_{\lvert\mathcal{G}}\Vert Q_{\lvert\mathcal{G}})\leq D_{\alpha
}(P\Vert Q)
\]
for any $\sigma$-subalgebra $\mathcal{G}\subseteq\mathcal{F}$, where
$P_{\lvert\mathcal{G}}$ and $Q_{\lvert\mathcal{G}}$ denote the restrictions of
$P$ and $Q$ to $\mathcal{G}$. As a special case, taking $\mathcal{G}%
=\{0,\mathcal{X}\}$ to be the trivial algebra, we find that
\[
D_{\alpha}(P\Vert Q)\geq0.
\]
$D_{\alpha}(P\Vert Q)=0$ if and only if $P=Q.$ Taking $\mathcal{G}%
=\sigma(\mathcal{P})$ to be the $\sigma$-algebra generated by a finite
partition $\mathcal{P}$ of $\mathcal{X}$, the data processing inequality
implies that discretizing $\mathcal{X}$ can only decrease $D_{\alpha}$.
However, because of the following property, which carries over from
$f$-divergences, $D_{\alpha}$ may be approximated arbitrarily well by such
finite partitions:
\begin{equation}
D_{\alpha}(P\Vert Q)=\sup_{\mathcal{P}}D_{\alpha}(P_{\lvert\sigma
(\mathcal{P})}\Vert Q_{\lvert\sigma(\mathcal{P})})\qquad(\alpha>0),
\label{eqn:supfinite}%
\end{equation}
where the supremum is over all finite partitions $\mathcal{P}$ of
$\mathcal{X}$. This characterization also shows that we have found the right
generalization of R\'{e}nyi's definition for finite $\mathcal{X}$.

Using the dominated convergence theorem it can be shown that:

\begin{theorem}
$D_{\alpha}$ is continuous in $\alpha$ on
\[
A=\{\alpha\mid0\leq\alpha\leq1\text{ or }D_{\alpha}(P\Vert Q)<\infty\}.
\]
$D_{\alpha}$ is also nondecreasing in $\alpha$, and on $A$ it is constant if
and only if $q/p$ is constant $P$-a.s.
\end{theorem}

The fact that $D_{\alpha}$ is nondecreasing, together with Equation
\eqref{eqn:supfinite}, implies that $\lim_{\alpha\uparrow1}D_{\alpha}=D$, as
asserted in our definition of $D_{1}$: for finite $\mathcal{X}$, this can be
verified directly using l'H\^{o}pital's rule. Therefore
\begin{multline}
 \lim_{\alpha \uparrow 1} D_\alpha(P\Vert Q) = \sup_{\alpha < 1} \sup_{\mathcal{P}} D_\alpha(P_{\lvert \sigma(\mathcal{P})} \Vert Q_{\lvert \sigma(\mathcal{P})}) \\ = \sup_{\mathcal{P}} \sup_{\alpha < 1} D_\alpha(P_{\lvert \sigma(\mathcal{P})}\Vert Q_{\lvert \sigma(\mathcal{P})}) = D(P\Vert Q). \label{eqn:d1}\end{multline} %
The assertion that $\lim_{\alpha\downarrow0}D_{\alpha}=-\log Q(p>0)$ is
verified differently, using the dominated convergence theorem and the
observation that $\lim_{\alpha\downarrow0}p^{\alpha}q^{1-\alpha}$ equals $q$
if $p>0$ and $0$ otherwise. R\'{e}nyi divergence may be extended to
$\alpha=\infty$ by letting $\alpha$ tend to $\infty$. Then, for finite
$\mathcal{X}$,
\[
D_{\infty}(P\Vert Q)=\log\max_{x\in\mathcal{X}}\frac{P(x)}{Q(x)},
\]
and by an interchanging of suprema similar to \eqref{eqn:d1} we find that
\[
D_{\infty}(P\Vert Q)=\log\sup_{A\in\mathcal{F}}\frac{P(A)}{Q(A)}%
=\log\esssup_{x \in\mathcal{X}}\frac{\mathrm{d}P}{\mathrm{d}Q}\left(
x\right)
\]
in general. Consequently, $D_{\infty}(Q\Vert P)$ (note the reversal of $P$ and
$Q$) is a one-to-one function of the separation distance $s(P,Q)=\max
_{x}(1-P(x)/Q(x))$, defined only for countable $\mathcal{X}$, which has been
used to obtain bounds on the rate of convergence to the stationary
distribution for certain Markov chains \cite{Gibbs2002,AldousDiaconis1987}.

Equation~\ref{eqn:supfinite} implies that there exists a sequence
$\mathcal{F}_{1},\mathcal{F}_{2},\ldots$ of $\sigma$-algebras generated by
finite partitions such that
\[
\lim_{n\rightarrow\infty}D_{\alpha}(P_{\lvert\mathcal{F}_{n}}\Vert
Q_{\lvert\mathcal{F}_{n}})=D_{\alpha}(P\Vert Q).
\]
By the connection to $f$-divergences, such a convergence result holds for any
\emph{increasing} sequence of $\sigma$-algebras $\mathcal{F}_{1}%
\subseteq\mathcal{F}_{2}\subseteq\cdots\subseteq\mathcal{F}_{\infty}%
=\sigma\left(  \bigcup_{n=1}^{\infty}\mathcal{F}_{n}\right)  \subseteq
\mathcal{F}$:%
\begin{equation}
\lim_{n\rightarrow\infty}D_{\alpha}(P_{\lvert\mathcal{F}_{n}}\Vert
Q_{\lvert\mathcal{F}_{n}})=D_{\alpha}(P_{\lvert\mathcal{F}_{\infty}}\Vert
Q_{\lvert\mathcal{F}_{\infty}})\qquad(\alpha>0)
\label{eqn:LimitAlgebraIncreasing}%
\end{equation}
\cite[Theorem 15]{Liese2006}. By a suitable choice of $\mathcal{F}_{n}$ this
result extends \emph{additivity} for any distributions $P_{1},P_{2},\ldots$
and $Q_{1},Q_{2},\ldots$,
\[
\sum_{n=1}^{N}D_{\alpha}(P_{n}\Vert Q_{n})=D_{\alpha}(P_{1}\times\cdots\times
P_{N}\Vert Q_{1}\times\cdots\times Q_{N}),
\]
from any finite $N$ (for which it is easy to prove) to $N=\infty$ (if
$\alpha>0$). For $\alpha=0$ additivity only holds for finite $N$. By a direct
proof we can also prove the counterpart to \eqref{eqn:LimitAlgebraIncreasing}
for decreasing sequences of $\sigma$-algebras $\mathcal{F}\supseteq
\mathcal{F}_{1}\supseteq\mathcal{F}_{2}\supseteq\cdots\supseteq\mathcal{F}%
_{\infty}=\bigcap_{n=1}^{\infty}\mathcal{F}_{n}$ (for finite $\alpha$) under
the condition that the divergence is finite.

Let
\[
H^{2}(P,Q)=\int(p^{1/2}-q^{1/2})^{2}\ \mathrm{d}\mu=2-2d_{1/2}(P,Q)
\]
denote the squared \emph{Hellinger distance}, and let
\[
\chi^{2}(P,Q)=\int\frac{(p-q)^{2}}{q}\mathrm{d}\mu=d_{2}(P,Q)-1
\]
denote the \emph{$\chi^{2}$-distance} \cite{Gibbs2002}. We see that
\[
D_{1/2}(P\Vert Q)=-2\log(1-H^{2}(P,Q)/2)
\]
and $D_{2}(P\Vert Q)=\log(1+\chi^{2}(P,Q))$. Hence by $\log x\leq x-1$
\begin{multline*}
H^{2}(P,Q)\leq D_{1/2}(P\Vert Q)\leq D(P\Vert Q)\\
\leq D_{2}(P\Vert Q)\leq\chi^{2}(P,Q).
\end{multline*}

\section{Majorization, Markov ordering and Lorenz
diagrams\label{sec:majorization}}

The general theory of majorization is now a well established mathematical
discipline \cite{Marshall1979}. The majorization lattice and its relation to
discrete entropy was studied in \cite{Cicalese2002} and later generalized in
\cite{Harremoes2004b}. Recently a long article on this subject by Gorban,
Gorban, and Judge has been accepted for publication \cite{Gorban2010}. We
refer to these papers for a more complete discussion and further references.
Here we shall relate the relative majorization lattice to R\'{e}nyi divergence.

\input{Lorenz.TpX}

\begin{definition}
Let $P$ and $Q$ be measures on the same measurable set. The \emph{Lorenz
diagram} of $\left(  P,Q\right)  $ is the range of
\[
f\mapsto\left(  \int f~\mathrm{d}P,\int f~\mathrm{d}Q\right)  ,
\]
where $f$ is any measurable function with values in $\left[  0,1\right]  .$
\end{definition}

If $Q$ is the uniform distribution then the Lorenz diagram of $\left(
P_{1},Q\right)  $ is a subset of the Lorenz diagram of $\left(  P_{2}%
,Q\right)  $ if and only if $P_{2}$ \emph{majorizes} $P_{1}.$

\begin{theorem}
The Lorenz diagram of $\left(  P_{1},Q\right)  $ is a subset of the Lorenz
diagram of $\left(  P_{2},Q\right)  $ if and only if there exists a Markov
operator that transforms $P_{2}$ into $P_{1}$ and leaves $Q$ invariant.
\end{theorem}

\begin{definition}
Let $P_{1},P_{2}$ and $Q$ be measures on the same measurable set $\mathcal{X}%
$. We write $P_{2}\succeq_{Q}P_{1}$ if the Lorenz diagram of
$\left(  P_{1},Q\right)  $ is a subset of the Lorenz diagram of $\left(
P_{2},Q\right)  .$ If the Lorenz diagrams of $(P_{1},Q)$ and $(P_{2},Q)$ are
equal, then we
write $P_{1}\simeq_{Q}P_{2}$.
\end{definition}

This ordering that generalizes majorization will be celled the \emph{Markov
ordering} \cite{Gorban2010}\footnote{In \cite{Harremoes2004b} this ordering
was called \emph{relative majorization}.}.

\begin{theorem}
[\cite{Harremoes2004b}]\label{lattice}Let $Q$ be a measure on a measurable set
$\mathcal{X}.$ If $Q$ is a uniform distribution on a finite set or if $Q$ has
no atoms, then $M_{+}^{1}\left(  \mathcal{X}\right)  /\simeq_{Q}$ is a
lattice, where $M_{+}^{1}\left(  \mathcal{X}\right)  $ denotes the set of
probability measures on $\mathcal{X}.$
\end{theorem}

The Lorenz diagram is characterized by a lower bound curve
that is convex and an upper bounding curve that is concave. Because of the
symmetry around $\left(  1/2,1/2\right)  $ the Lorenz diagram is completely
determined by the lower bounding curve.

\begin{definition}
The \emph{Lorenz curve} of $\left(  P,Q\right)  $ is the convex envelope of
the Lorenz diagram, i.e.\ the largest convex function such that all the points
in the Lorenz diagram are at or above the curve.
\end{definition}

\begin{proposition}
[\cite{Harremoes2004b}]Let $P$ and $Q$ be measures on the same measurable set
$\mathcal{X}$. The Lorenz curve of $\left(  P,Q\right)  $ is the convex
envelop of the points $\left(  P\left(  A_{t}\right)  ,Q\left(  A_{t}\right)
\right)  $ where $A_{t}$ are events of the form $A_{t}=\left\{  x\in
\mathcal{X}\mid\frac{\mathrm{d}P}{\mathrm{d}Q}\leq t\right\}  $.
\end{proposition}

In statistics the sets $A_{t}=\left\{  x\in\mathcal{X}\mid\frac{\mathrm{d}%
P}{\mathrm{d}Q}\leq t\right\}  $ play the role of acceptance sets related to
the likelihood ratio test of ratio $t$. The proof of this proposition is
therefore essentially the same as the proof of the Neyman-Pearson Lemma
\cite{Harremoes2004b}. Note that for discrete measures there will only be
finitely many different points of the form $\left(  P\left(  A_{t}\right)
,Q\left(  A_{t}\right)  \right)  ,$ and in that case the Lorenz curve is
piecewise linear. For $t_{1}<t_{2}$%
\[
\frac{P\left(  A_{t_{2}}\right)  -P\left(  A_{t_{1}}\right)  }{Q\left(
A_{t_{2}}\right)  -Q\left(  A_{t_{1}}\right)  }=\frac{P\left(  \left\{  x\mid
t_{1}<\frac{\mathrm{d}P}{\mathrm{d}Q}\leq t_{2}\right\}  \right)  }{Q\left(
\left\{  x\mid t_{1}<\frac{\mathrm{d}P}{\mathrm{d}Q}\leq t_{2}\right\}
\right)  }\in\left]  t_{1},t_{2}\right]  ,
\]
so $\left(  P\left(
A_{t}\right)  ,Q\left(  A_{t}\right)  \right)  $ gives a parametrization of
the Lorenz curve in terms of its slope if it is differentiable.

Suppose $Q$ is the counting measure on a finite set $\mathcal{X}$ of size $n$,
and let $P_{1}=(v_{1},\ldots,v_{n})$ be a discrete measure on $\mathcal{X}$.
Then $A_{t}$ is simply $\left\{  i\mid v_{i}\leq t\right\}  .$ Let
$P_{2}=(w_{1},\ldots,w_{n})$ be another measure and let $B_{t}=\left\{  i\mid
w_{i}\leq t\right\}  $. Then $P_{1}\preceq P_{2}$ if and only if $P_{1}\left(
A_{t_{1}}\right)  \geq P_{2}\left(  B_{t_{2}}\right)  $ whenever $Q\left(
A_{t_{1}}\right)  =Q\left(  B_{t_{2}}\right)  $. Thus $P_{1}\preceq P_{2}$ if
and only if the Lorenz curve of $\left(  P_{1},Q\right)  $ is above the Lorenz
curve of $\left(  P_{2},Q\right)  $.

If one of the conditions of Theorem \ref{lattice} is fulfilled, then for each
convex function $f$ there exists a measure $P$ such that $f$ is the Lorenz
curve of $P$. Thus $M_{+}^{1}\left(  \mathcal{X}\right)  /\simeq_{Q}$ can be
identified with the set of Lorenz curves. Let $P_{1}$ and $P_{2}$ be measures
and let $L_{1}$ and $L_{2}$ be their Lorenz curves. Then $P_{1}\wedge P_{2}$
can be identified with the Lorenz curve $\max\left\{  L_{1},L_{2}\right\}  $
and $P_{1}\vee P_{2}$ can be identified with the Lorenz curve that is the
convex envelop of $\min\left\{  L_{1},L_{2}\right\}  $. In general this
lattice is neither modular nor distributive \cite{Harremoes2004b}.\input{metjoin.TpX}

\section{Divergence, Convexity and Ordering\label{sec:DivergenceMajorization}}

We will now consider properties of $D_{\alpha}(P\Vert Q)$ as we vary $P$ and
$Q$ while keeping $\alpha$ fixed. Information divergence $D(P\Vert Q)$ is known to be jointly \emph{convex} in
the pair $(P,Q)$ \cite{Cover1991}. By an argument similar to the proof for
$D_{1}$ in \cite{Cover1991}, this property generalizes to $D_{\alpha}$ for
arbitrary order $0\leq\alpha\leq1$:

\begin{theorem}
For $0\leq\alpha\leq1$, $D_{\alpha}(P,Q)$ is jointly convex in the pair
$(P,Q)$.
\end{theorem}

Even though joint convexity does not generalize to $\alpha>1$, we still have:

\begin{theorem}
For all $\alpha$, $D_{\alpha}(P\Vert Q)$ is convex in $Q.$
\end{theorem}

The key step in proving the latter result for $\alpha>1$ relies on
H\"{o}lder's inequality.

Let $P$ be absolutely continuous with respect to $Q$. If $F$ denotes the curve
that upper bounds the Lorenz diagram, then the R\'{e}nyi divergence is given
by
\[
D_{\alpha}\left(  P\Vert Q\right)  =\frac{1}{\alpha-1}\log\int_{0}^{1}\left(
F^{\prime}\left(  t\right)  \right)  ^{\alpha}~\mathrm{d}t.
\]
Note that we can replace the upper bounding function by the lower bounding
function (the Lorenz curve) without changing the integral.

\begin{theorem}
\label{decreasing}For $\alpha>0$ the R\'{e}nyi divergence $D_{\alpha}\left(
P\Vert Q\right)  $ is a increasing function of $P$ on the lattice
corresponding to $Q$.
\end{theorem}

\begin{IEEEproof}
Let $F$ and $G$ be concave functions on $\left[  0,1\right]  $ such that
$F\leq G$ and $F\left(  0\right)  =G\left(  0\right)  =0$ and $F\left(
1\right)  =G\left(  1\right)  =1.$ Let $x\mapsto\Phi_{x}$ be a Markov kernel
such that $x=\int y~\mathrm{d}\Phi_{x}(y)$ for all $x\in\left[  0,\infty
\right[  $. Then
\begin{equation}
\int G(y)\ \mathrm{d}\Phi_{x}(y)\leq G\left(  \int y\ \mathrm{d}\Phi
_{x}(y)\right)  =G\left(  x\right)  .
\end{equation}
Consider the set of all Markov kernels $x\mapsto\Phi_{x}$ such that $x=\int
y\ \mathrm{d}\Phi_{x}(y)$ and $F\left(  x\right)  \leq\int G\left(  y\right)
\ \mathrm{d}\Phi_{x}(y)$ for all $x\in\left[  0;\infty\right[  $. This set is
convex and contains an element such that $F\left(  x\right)  =\int G\left(
y\right)  \ \mathrm{d}\Phi_{x}(y).$ Then $F^{\prime}\left(  x\right)  =\int
G^{\prime}\left(  y\right)  \ \mathrm{d}\Phi_{x}(y)$ and the theorem follows
from Jensen's inequality.
\end{IEEEproof}

Theorem \ref{decreasing} is essentially a noisy data processing inequality
because the Markov kernel $\Phi_{x}$ in the proof essentially maps the measure
corresponding to $G$ into the measure corresponding to $P.$ By adapting a proof from \cite{Harremoes2004b} is possible to prove the following theorem:

\begin{theorem}
\label{subsuper}Let $P_{1}$ and $P_{2}$ denote distributions that are
absolutely continuous with respect to $Q$. If Markov ordering is taken with
respect to $Q$ then power divergence is sub-modular and super-additive, i.e.

\begin{align*}
d_{\alpha}\left(  P_{1},Q\right)  +d_{\alpha}\left(  P_{2},Q\right) &\geq d_{\alpha}\left(  P_{1}\wedge P_{2},Q\right)  +d_{\alpha}\left(
P_{1}\vee P_{2},Q\right)\\
\intertext{and}
d_{\alpha}\left(  P_{1},Q\right)  +d_{\alpha}\left(  P_{2},Q\right)  &\leq d_{\alpha}\left(  P_{1}\wedge P_{2},Q\right)  .
\end{align*}

\end{theorem}

Since power divergence is a function of R\'{e}nyi divergence one can
reformulate Theorem \ref{subsuper} in terms of R\'{e}nyi divergence. Like R\'{e}nyi divergence, the power divergence $d_{\alpha}(P,Q)$ tends to the
information divergence $D(P\Vert Q)$ as $\alpha\uparrow1$. This
implies:

\begin{corollary}
Let $P_{1}$ and $P_{2}$ be distributions that are absolutely continuous with
respect to $Q$. If the Markov ordering is taken with respect to $Q$ then
information divergence is sub-modular and super-additive, i.e.%
\begin{align*}
D\left(  P_{1}\Vert Q\right)  +D\left(  P_{2}\Vert Q\right)   &  \geq D\left(
P_{1}\wedge P_{2}\Vert Q\right)  +D\left(  P_{1}\vee P_{2}\Vert Q\right)  \\
D\left(  P_{1}\Vert Q\right)  +D\left(  P_{2}\Vert Q\right)   &  \leq D\left(
P_{1}\wedge P_{2}\Vert Q\right)  .
\end{align*}

\end{corollary}

\section{Continuity of R\'{e}nyi divergence\label{sec:continuity}}

The type of continuity of $D_{\alpha}$ in the pair $(P,Q)$ turns out to depend
on the topology and on $\alpha$. We consider the \emph{$\tau$-topology}, in
which convergence of $P_{n}$ to $P$ means that $P_{n}(A)\rightarrow P(A)$ for
all $A\in\mathcal{F}$, and the \emph{total variation topology} in which
$P_{n}\rightarrow P$ if the variation distance between $P_{n}$ and $P$ goes to
zero. In general the total variation topology is stronger than the $\tau
$-topology, but if $\mathcal{X}$ is countable, then the two topologies coincide.

\begin{theorem}
For any $\alpha> 0$, $D_{\alpha}(P\Vert Q)$ is a lower semi-continuous
function of $(P,Q)$ in the $\tau$-topology.
\end{theorem}

Moreover:

\begin{theorem}
For $0 < \alpha< 1$, $D_{\alpha}(P \Vert Q)$ is a (uniformly) continuous
function of $(P,Q)$ in the total variation topology.
\end{theorem}

It remains to consider $\alpha= 0$. In this case:

\begin{corollary}
$D_{0}(P\Vert Q)$ is an upper semi-continuous function of $(P,Q)$ in the total
variation topology.
\end{corollary}

Using the Markov ordering we get more insight.

\begin{theorem}
If $\alpha\geq1$ and $D_{\alpha}(\tilde{P}\Vert Q)<\infty$, then the R\'{e}nyi
divergence $D_{\alpha}\left(  P\Vert Q\right)  $ is continuous in $P$ on the
set $\left\{  P\mid P\preceq_{Q}\tilde{P}\right\}  $ when the set of
probability measures is equipped with the topology of total variation.
\end{theorem}

\begin{IEEEproof}
If $P_{n}\rightarrow P$ in total variation for $n\rightarrow\infty$ then the
Lorenz diagram of $P_{n}$ tends to the Lorenz diagram of $P$ in Hausdorff
distance. Let $F,\tilde{F}$ and $F_{n}$ denote the upper bounding functions
for $P,\tilde{P}$ and $P_{n}$. Then for any $\varepsilon>0$ eventually
$F_{n}\left(  t\right)  \leq\min\left\{  \tilde{F}\left(  t\right)  ,F\left(
t+\varepsilon\right)  \right\}  $ for all $t\in\left[  0,1\right]  .$ Hence
\begin{multline*}
\lim\sup_{n \to\infty} D_{\alpha}\left(  P_{n}\Vert Q\right) \\
\leq\frac{1}{\alpha-1}\log\int_{0}^{1}\left(  \frac{d}{dt}\min\left\{
\tilde{F}\left(  t\right)  ,F\left(  t+\varepsilon\right)  \right\}  \right)
^{\alpha}~dt.
\end{multline*}
This holds for all $\varepsilon>0$ and, since the right-hand side tends to
$\frac{1}{\alpha-1}\log\int_{0}^{1}\left(  \frac{d}{dt}F\left(  t\right)
\right)  ^{\alpha}~dt=D_{\alpha}\left(  P\Vert Q\right)  $ for $\varepsilon
\rightarrow0$, the result follows.
\end{IEEEproof}

\section{Guessing moments\label{sec:guessing}}

Erdal Arikan observed that the discrete version of R\'{e}nyi entropy is
related to so-called guessing moments \cite{Arikan1996}. In this short note we
shall see that R\'{e}nyi divergences are also related to guessing moments.

\begin{definition}
Let $P_{1}$ and $P_{2}$ denote probability measures on $\mathcal{X}.$ We say
that $P_{1}$ is \emph{a rearrangement} of $P_{2}$ if
\[
Q\left\{  x\in\mathcal{X}\mid\frac{\mathrm{d}P_{1}}{\mathrm{d}Q}\left(
x\right)  \geq t\right\}  =Q\left\{  x\in\mathcal{X}\mid\frac{\mathrm{d}P_{2}%
}{\mathrm{d}Q}\left(  x\right)  \geq t\right\}
\]
for all $t\in\mathbb{R}.$
\end{definition}

\begin{definition}
\emph{A guessing function} in $\mathcal{X}$ is a function $g:\mathcal{X}%
\rightarrow\mathbb{R}$ such that $Q\left(  \left\{  x\mid g\left(  x\right)
\leq t\right\}  \right)  \leq t$ for $t\in\left[  0,1\right]  .$
\end{definition}

For a probability measure $P$ on $\mathcal{X}$ with density $\frac
{\mathrm{d}P}{\mathrm{d}Q}$ we are interested in bounds on the moments of
guessing functions. For a guessing function $g$ the $\rho$-th moment is given
by%
\[
\left\Vert g\right\Vert _{\rho}=\left(  \int_{\mathbb{R}^{d}}\left(  g\left(
x\right)  \right)  ^{\rho}\ \mathrm{d}P(x)\right)  ^{1/\rho}.
\]

\begin{definition}
Let $P$ be a probability measure on $\mathcal{X}.$ For each Radon-Nikod\'ym
derivative $\frac{\mathrm{d}P}{\mathrm{d}Q}$, the \emph{ranking function} $r$
of $\frac{\mathrm{d}P}{\mathrm{d}Q}$ is given by%
\[
r\left(  x\right)  =Q\left(  \left\{  y\mid\frac{\mathrm{d}P}{\mathrm{d}%
Q}\left(  y\right)  \geq\frac{\mathrm{d}P}{\mathrm{d}Q}\left(  x\right)
\right\}  \right)  .
\]

\end{definition}

We note that if $F$ is the distribution function of $\frac{\mathrm{d}%
P}{\mathrm{d}Q}$ then the ranking function is given by $r\left(  x\right)
=1-F\left(  x\right)  $. The ranking function is a guessing function.%
\begin{multline*}
Q\left(  \left\{  x\mid r\left(  x\right)  \leq t\right\}  \right)  =\\
Q\left(  \left\{  x\mid Q\left(  \left\{  y\mid\frac{\mathrm{d}P}{\mathrm{d}%
Q}\left(  y\right)  \geq\frac{\mathrm{d}P}{\mathrm{d}Q}\left(  x\right)
\right\}  \right)  \leq t\right\}  \right)  \leq t.
\end{multline*}
Note that $Q\left(  \left\{  x\mid r\left(  x\right)  \leq t\right\}  \right)
=t$ for all $t\in\left[  0,1\right]  $ if and only if the distribution of the
random variable $\frac{\mathrm{d}P}{\mathrm{d}Q}$ is continuous.

\begin{proposition}
The ranking function is the guessing function that minimizes the $\rho$-th
moment if $\rho>0$ and maximizes the $\rho$-th moment if $\rho<0$.
\end{proposition}

Guessing and ranking are closely related to majorization and the Markov
ordering via the following proposition.

\begin{proposition}
Assume that $P_{1},P_{2}$ and $Q$ are probability measures on $\mathcal{X}$
and $P_{1}\preceq_{Q}P_{2}.$ Let $r_{1}$ and $r_{2}$ denote the ranking
functions of $P_{1}$ and $P_{2}.$ Then
\begin{align*}
\left\Vert r_{1}\right\Vert _{\rho}  &  \leq\left\Vert r_{2}\right\Vert
_{\rho}\text{ if }\rho>0,\\
\left\Vert r_{1}\right\Vert _{\rho}  &  \geq\left\Vert r_{2}\right\Vert
_{\rho}\text{ if }\rho<0.
\end{align*}

\end{proposition}

\begin{lemma}
\label{nedre}If $\alpha=\frac{1}{1+\rho}>0$ then, for any probability measures
$P$ and $Q$,
\[
-\log\left(  \left\Vert r\right\Vert _{\rho}\right)  \geq D_{\alpha}\left(
P\Vert Q\right)  ,
\]
where the $\rho$-norm is calculated with respect to $Q$ and $r$ is the ranking
function of $\frac{\mathrm{d}P}{\mathrm{d}Q}.$
\end{lemma}

\begin{IEEEproof}
We have
\begin{align*}
r\left(  x\right)   &  =\int_{\frac{\mathrm{d}P}{\mathrm{d}Q}\left(  y\right)
\geq\frac{\mathrm{d}P}{\mathrm{d}Q}\left(  x\right)  }1~\mathrm{d}%
Q(y)=\int_{\frac{\mathrm{d}P}{\mathrm{d}Q}\left(  y\right)  \geq
\frac{\mathrm{d}P}{\mathrm{d}Q}\left(  x\right)  }1^{\alpha}~\mathrm{d}Q(y)\\
&  \leq\int_{\frac{\mathrm{d}P}{\mathrm{d}Q}\left(  y\right)  \geq
\frac{\mathrm{d}P}{\mathrm{d}Q}\left(  x\right)  }\left(  \frac{\frac
{\mathrm{d}P}{\mathrm{d}Q}\left(  y\right)  }{\frac{\mathrm{d}P}{\mathrm{d}%
Q}\left(  x\right)  }\right)  ^{\alpha}~\mathrm{d}Q(y)\\
&  \leq\int\left(  \frac{\frac{\mathrm{d}P}{\mathrm{d}Q}\left(  y\right)
}{\frac{\mathrm{d}P}{\mathrm{d}Q}\left(  x\right)  }\right)  ^{\alpha
}~\mathrm{d}Q(y)=\frac{\int\left(  \frac{\mathrm{d}P}{\mathrm{d}Q}\left(
y\right)  \right)  ^{\alpha}~\mathrm{d}Q(y)}{\left(  \frac{\mathrm{d}%
P}{\mathrm{d}Q}\left(  x\right)  \right)  ^{\alpha}}.
\end{align*}
We get%
\begin{multline*}
E\left[  r\left(  X\right)  ^{\rho}\right]  \leq\int\left(  \frac{\int\left(
\frac{\mathrm{d}P}{\mathrm{d}Q}\left(  y\right)  \right)  ^{\alpha}%
~\mathrm{d}Q(y)}{\left(  \frac{\mathrm{d}P}{\mathrm{d}Q}\left(  x\right)
\right)  ^{\alpha}}\right)  ^{\rho}~\frac{\mathrm{d}P}{\mathrm{d}Q}\left(
x\right)  ~\mathrm{d}Q(x)\\
=\left(  \int\left(  \frac{\mathrm{d}P}{\mathrm{d}Q}\left(  y\right)  \right)
^{\alpha}~\mathrm{d}Q(y)\right)  ^{\rho}\int\left(  \frac{\mathrm{d}%
P}{\mathrm{d}Q}\left(  x\right)  \right)  ^{1-\alpha\rho}~\mathrm{d}Q(x)\\
=\left(  \int~\left(  \frac{\mathrm{d}P}{\mathrm{d}Q}\left(  x\right)
\right)  ^{\alpha}~\mathrm{d}Q(x)\right)  ^{\frac{1}{\alpha}}.
\end{multline*}
We raise to the power $1/\rho$ and take minus the logarithm and get%
\begin{multline*}
\log\left(  E\left[  r\left(  X\right)  ^{\rho}\right]  ^{\frac{1}{\rho}}\right)  
\leq\log\left(  \left(  \int~\left(  \frac{\mathrm{d}P}{\mathrm{d}Q}\left(
x\right)  \right)  ^{\alpha}~\mathrm{d}Q(x)\right)  ^{\frac{1}{\alpha\rho}%
}\right)  \\
=\frac{1}{1-\alpha}\log\left(  \int~\left(  \frac{\mathrm{d}P}{\mathrm{d}%
Q}\left(  x\right)  \right)  ^{\alpha}~\mathrm{d}Q(x)\right)  =-D_{\alpha
}\left(  P\Vert Q\right)  .
\end{multline*}

\end{IEEEproof}

Using additivity of R\'{e}nyi divergence and Lemma \ref{nedre} we get the
following theorem.

\begin{theorem}
If $\alpha=\frac{1}{1+\rho}>0$ then for any i.i.d.\ sequence $X_{1}%
^{n}=\left(  X_{1},X_{2},\ldots,X_{n}\right)  \in\mathcal{X}^{n}$ we have
\[
-\frac{1}{n}\log\left(  \left\Vert r\left(  X_{1}^{n}\right)  \right\Vert
_{\rho}\right)  \geq D_{\alpha}\left(  P\Vert Q\right)  .
\]
This bound is asymptotically tight as stated in the following theorem.
\end{theorem}

\begin{theorem}
If $\alpha=\frac{1}{1+\rho}>0$ then for any i.i.d.\ sequence $X_{1}%
^{n}=\left(  X_{1},X_{2},\ldots,X_{n}\right)  \in\mathcal{X}^{n}$ we have
\[
\lim_{n\rightarrow\infty}-\frac{1}{n}\log\left(  \left\Vert r\left(  X_{1}%
^{n}\right)  \right\Vert _{\rho}\right)  =D_{\alpha}\left(  P\Vert Q\right)
.
\]

\end{theorem}

The result gives a new interpretation of R\'{e}nyi divergence.

\section{Discussion}

The results in this short paper are formulated under the assumption that the
second argument $Q$ in $D_{\alpha}\left(  P\Vert Q\right)  $ is a probability
measure. Nevertheless many of the results still hold if $Q$ is a more general
positive measure. For instance many results on R\'{e}nyi entropy are obtained
when $Q$ denotes the counting measure. Most of these results for R\'{e}nyi
entropy are well-known. Results for differential R\'{e}nyi entropy are
obtained when $Q$ is the Lebesgue measure. For both R\'{e}nyi entropy and
differential R\'{e}nyi entropy many results should first be formulated and
proved for subsets of finite measure and then one should take a limit for an
increasing sequence of subsets. In this sense our results on R\'{e}nyi
divergence are often more general than the results one will find in the
literature. 

We have related R\'{e}nyi divergence to majorization and Markov
ordering. An interesting related concept is catalytic majorization. It has
been proved by M. Klimesh that one discrete distribution majorizes another
distribution if and only if certain inequalities hold between their R\'{e}nyi
entropies \cite{Klimesh2004}. A similar result is still to be proved for
R\'{e}nyi divergence.

\section{Acknowledgments}

We thank Christophe Vignat, Matthew Klimesh, and Erdal Arikan for useful discussions.

This work was supported in part by the IST Programme of the European
Community, under the PASCAL Network of Excellence, IST-2002-506778.

\setlength{\itemsep}{5pt}

\bibliographystyle{ieeetr}
\bibliography{database1}

\end{document}